# Physiologically-Based Pharmacokinetic Modeling of Blood Clearance of Liver Fluorescent Markers for the Assessment of the Degree of Hepatic Ischemia-Reperfusion Injury*

Christopher Monti, Said H. Audi, PhD, Justin Womack, Seung-Keun Hong, PhD, Yongqiang Yang, PhD, Joohyun Kim, MD, PhD, Ranjan K. Dash, PhD, *Member, IEEE EMBS*

*Abstract*— During liver transplantation, ischemia-reperfusion injury (IRI) is inevitable and decreases the overall success of the surgery. While guidelines exist, there is no reliable way to quantitatively assess the degree of IRI present in the liver. Our recent study has shown a correlation between the bile-to-plasma ratio of FDA-approved sodium fluorescein (SF) and the degree of hepatic IRI, presumably due to IRI-induced decrease in the activity of the hepatic multidrug resistance-associated protein 2 (MRP2); however, the contribution of SF blood clearance via the bile is still convoluted with other factors, such as renal clearance. In this work, we sought to computationally model SF blood clearance via the bile. First, we converted extant SF fluorescence data from rat whole blood, plasma, and bile to concentrations using calibration curves. Next, based on these SF concentration data, we generated a "liver-centric", physiologically-based pharmacokinetic (PBPK) model of SF liver uptake and clearance via the bile. Model simulations show that SF bile concentration is highly sensitive to a change in the activity of hepatic MPR2. These simulations suggest that SF bile clearance along with the PBPK model can be used to quantify the effect of IRI on the activity of MRP2.

*Clinical Relevance*— This study establishes the theory necessary to generate a model for predicting the degree of IRI during liver transplantation.

## I. INTRODUCTION

Hepatic ischemia-reperfusion injury (IRI) inherent to liver transplantation affects physiological processes including bile formation [1, 2]. Diminished bile formation has been implicated as a marker of liver injury; however, the kinetics of biological markers through the bile formation machinery have yet to provide a reliable diagnostic tool to predict liver viability after transplantation [3, 4]. In a recent study, we noted this lack of tool to evaluate the effects of liver IRI and developed a method to assess biliary function using an FDA-approved fluorescent dye, sodium fluorescein (SF) [5]. This assessment method involved determination of the bile-to-plasma ratio (BPR) of SF fluorescence over time, where lower BPR correlated with worse IRI. We suggested that the intracellular sequestration of the multi-drug resistance-associated protein 2 (MRP2) transporter during IRI, leading to an overall decrease in SF transport efficiency, was the causative factor for the lower BPR [5].

Given the complex, multifactorial nature of *in vivo* hepatic physiology, however, it is unclear how much SF is cleared by the active MRP2 transporters in relation to other factors such as the presence or absence of IRI, renal clearance, or distribution into other body compartments. Physiologically-based pharmacokinetic (PBPK) modeling provides an integrative and mechanistic framework allowing for the analysis and quantification of different vascular, tissue, and cellular processes contributing to the clearance of SF under physiological conditions and in response to IRI. In this manuscript, we *hypothesize* that MRP2 is an important contributor to SF disposition in the bile and that its diminishment will markedly decrease the amount of SF in the bile relative to other contributors to SF uptake and clearance by the liver. We tested this hypothesis by computationally modeling SF disposition in the bile using a "liver-centric" PBPK model, and by simulating the effects of a change in each of the model parameters descriptive of the dominant tissue processes that determine SF clearance through the liver, including MRP2, on SF blood, plasma, and bile concentrations.

## II. METHODS

*A. Fluorescence to Concentration Calibration Experiments*

SF fluorescence data were obtained from our previous work [5]. While SF fluorescence ratio measurements (i.e., BPR) are sufficient for qualitative analysis of hepatic IRI, measurements of SF concentrations are needed for PBPK modeling.

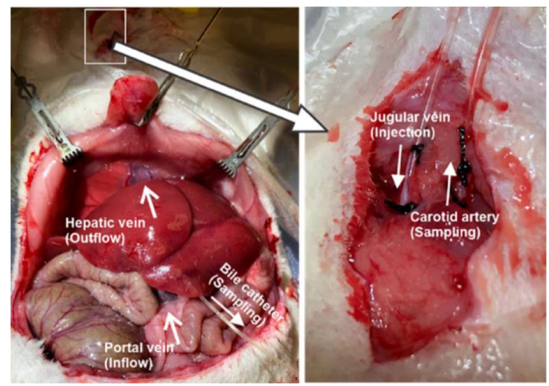

**Fig. 1:** Photographs of *in vivo* experimental procedure showing pertinent anatomy and injection and sampling sites.

*Research partly supported by the Joint MU-MCW Department of BME Pilot Product Development Grant FP00022381 to R.K.D. and The Kevin T. Cottrell Memorial Fund for Organ Transplantation and The J. Scott Harkness Organ Transplantation Research and Education Fund to J.K.

All Authors except SA are with the Medical College of Wisconsin (cmonti@mcw.edu, jwomack@mcw.edu, skhong@mcw.edu, yoyang@mcw.edu, jokim@mcw.edu, rdash@mcw.edu; (414) 955-4497). SA is with Marquette University (said.audi@marquette.edu).

Thus, we performed calibration experiments and developed calibration curves to relate SF fluorescence measurements to known concentrations of SF in blood, plasma, and bile. Using Sprague-Dawley (SD) rats (Charles River Laboratories, Chicago, IL), we obtained blood, plasma, and bile samples (Fig. 1) [5]. We then spiked in known concentrations of SF (Sigma-Aldrich, St. Louis, MO) in 0.9% saline and measured fluorescence using a CLARIOStar Microplate Reader (BMG LABTECH) with the gain set at 1,000 for blood, 600 for plasma, and 500 for bile [5]. The blood and plasma calibration data were both obtained using 3 biological replicates and 8 technical replicates. The bile calibration data were obtained using 2 biological replicates with 7 technical replicates. Empirical equations (1)-(3) were fitted to the calibration data using linear and nonlinear regression. Parameter values from the regression analysis are listed in Table I.

$$I_{Bl} = Int + Sl \times C_{Bl} \tag{1}$$

$$I_{Pl} = \frac{I_{max,Pl} C_{Pl}}{C_{50,Pl} + C_{Pl}} \tag{2}$$

$$I_B = \frac{I_{max,B} C_B}{C_{50,B} + C_B} \tag{3}$$

Where, $I_i$ is the fluorescence intensity of SF, $C_i$ is the concentration of SF in a specific physiologic region, $I_{max,i}$ is the maximal fluorescence intensity, and $C_{50,i}$ is the SF concentration required to achieve 50% maximal intensity. Finally, using these calibration curves, we converted the SF fluorescence data from [5] to concentration.

### B. Empirical Modeling of Hepatic SF Input Concentration

To provide SF inflow for the governing ordinary differential equations (ODEs) of our "liver-centric" PBPK model for SF blood clearance through hepatocytes into bile (see section *C*), we formulated (4) and fitted this empirical equation to the blood SF concentration data using the MATLAB *fmincon* function. All physiologic parameter values used below are from [6] unless another reference is specified.

$$C_{in,Bl,SF}(t) = \begin{cases} (C_{peak}/t_{peak})t & 0 \leq t \leq t_{peak} \\ A_1 e^{-k_1 t} + A_2 e^{-k_2 t} + A_3 e^{-k_3 t} & t \geq t_{peak} \end{cases} \tag{4}$$

Where $C_{peak}$ is the maximum blood SF concentration after injection accounting for transit time through the system and is defined as the SF dose (2 mg/kg) multiplied by the average rat weight (0.277 kg) divided by the total blood volume (TBV, 20.70 mL). $t_{peak}$ is the time taken to achieve $C_{peak}$ and is defined as the quotient of the TBV and the cardiac output (CO, 85.05 mL/min). To ensure a continuous function, we imposed the parameter constraint that at $t_{peak}$, both pieces of the piecewise function (4) must be equal.

In blood (inflow/outflow), SF is distributed between plasma and red blood cells (RBCs) characterized by a variable partition coefficient ($\lambda$) that depends on the blood SF concentration. We empirically determined $\lambda$ in (5) to relate plasma SF concentration to blood SF concentration, given by (4) for inflow blood, and fitted the resultant curve to the inflow plasma SF concentration data using the relationship in (6). Using (7), we computed the SF RBC concentration.

$$\lambda = \frac{C_{in,Pl,SF}}{C_{in,RBC,SF}} = 0.5 + \frac{\lambda_{max}(C_{in,Bl,SF})^{n_\lambda}}{K_\lambda^{n_\lambda} + (C_{in,Bl,SF})^{n_\lambda}} \tag{5}$$

$$C_{in,Pl,SF} = \left(\frac{V_{RBC} + V_{Pl}}{V_{Pl} + \frac{V_{RBC}}{\lambda}}\right) C_{in,Bl,SF} \tag{6}$$

$$C_{in,RBC,SF} = \left(\frac{V_{RBC} + V_{Pl}}{V_{RBC} + \lambda V_{Pl}}\right) C_{in,Bl,SF} \tag{7}$$

Where $V_{RBC}$ is the volume of RBCs and is equal to the blood hematocrit (0.4) multiplied by the volume of blood in the liver (2.01 mL) and $V_{Pl}$ is the volume of plasma equal to 1- hematocrit (0.6) multiplied by the volume of blood in the liver. $\lambda_{max}$, $n_\lambda$, and $K_\lambda$ represent empirical parameters describing the concentration-dependent relationship between the plasma and RBC SF concentrations.

Finally, SF is metabolized to SF glucuronide (SFG) in the liver [7]. To account for this process, we used the human data from [8] demonstrating the concentration fractions of SF and SFG ($f_{SF}$ and $f_{SFG}$) in plasma over time, and scaled this time course data to account for differences in the vascular transit times (tt) between rats and humans (tt$_{rat}$ ≈ tt$_{human}$/4). We then formulated (8) and fitted this empirical equation to the time course data for $f_{SFG}$ to estimate the plasma concentration of SFG (and in the blood by rearranging (6)) given the concentration of SF in the plasma and the relative fraction of SFG.

$$f_{SFG} = \frac{f_{max,SFG} \, t^{n_{SFG}}}{t_{50,SFG}^{n_{SFG}} + t^{n_{SFG}}} \tag{8}$$

Where $t$ is the time measured in minutes after SF administration and $f_{SFG}$ is the fraction of SFG in plasma. $f_{max,SFG}$ is the maximum fraction of SFG achieved in the plasma, while $t_{50,SFG}$ is the time at which 50% SFG is achieved. Values for all parameters in this section can be found in Table II.

Equations (5)-(7) were applied to solutions of (9)-(10) to obtain outflow plasma and RBC SF and SFG concentrations.

### C. PBPK Modeling of SF Blood Clearance Via the Liver

Using the results from section *B* as inputs and anatomical guidance from Fig. 1, we developed an PBPK "liver-centric" model of SF and SFG blood clearance through hepatocytes into the bile as schematized in Fig. 2 and described mathematically in (9)-(14).

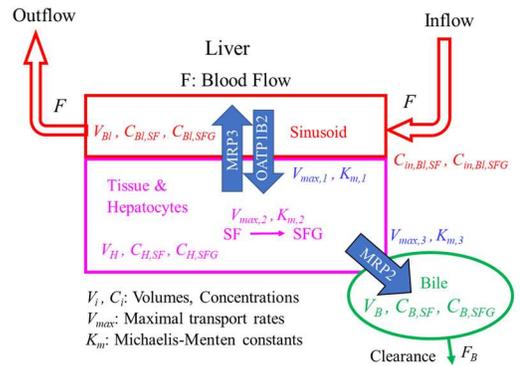

**Fig. 2:** Schematic of "liver-centric" PBPK model showing the processes of SF transport, metabolism, and clearance. Inflow represents blood coming into the liver from the hepatic artery and portal vein. Outflow represents blood leaving the liver via the hepatic vein. SF is converted to SFG in hepatocytes, and both are transported back to blood via OATP1B2/MRP3 and excreted to bile via MRP2. Variables and parameters are as defined in the text.

$$V_{Bl} \frac{dC_{Bl,SF}}{dt} = F(C_{in,Bl,SF} - C_{Bl,SF}) - \frac{V_{max,1}(C_{Bl,SF} - C_{H,SF})}{K_{m,1} + C_{Bl,SF} + C_{H,SF}} \tag{9}$$

$$V_{Bl}\frac{dC_{Bl,SFG}}{dt} = F(C_{in,Bl,SFG} - C_{Bl,SFG}) - \frac{V_{max,1}(C_{Bl,SFG} - C_{H,SFG})}{K_{m,1} + C_{Bl,SFG} + C_{H,SFG}} \quad (10)$$

$$V_H\frac{dC_{H,SF}}{dt} = \frac{V_{max,1}(C_{Bl,SF} - C_{H,SF})}{K_{m,1} + C_{Bl,SF} + C_{H,SF}} - \frac{V_{max,2}C_{H,SF}}{K_{m,2} + C_{H,SF}} - \frac{V_{max,3}C_{H,SF}}{K_{m,3} + C_{H,SF}} \quad (11)$$

$$V_H\frac{dC_{H,SFG}}{dt} = \frac{V_{max,1}(C_{Bl,SFG} - C_{H,SFG})}{K_{m,1} + C_{Bl,SFG} + C_{H,SFG}} + \frac{V_{max,2}\ C_{H,SF}}{K_{m,2} + C_{H,SF}} - \frac{V_{max,3}\ C_{H,SFG}}{K_{m,3} + C_{H,SFG}} \quad (12)$$

$$V_B\frac{dC_{B,SF}}{dt} = \frac{V_{max,3}\ C_{H,SF}}{K_{m,3} + C_{H,SF}} - F_B C_{B,SF} \quad (13)$$

$$V_B\frac{dC_{B,SFG}}{dt} = \frac{V_{max,3}\ C_{H,SFG}}{K_{m,3} + C_{H,SFG}} - F_B C_{B,SFG} \quad (14)$$

In this model, SF and SFG blood *inflow* concentrations to the liver ($C_{in,Bl,SF}$ and $C_{in,Bl,SFG}$) are as described above. The flow ($F$) into the liver is the sum of blood flow from the hepatic artery and portal vein, which, after intravenous administration, contain SF and SFG. In the liver, the blood from the hepatic artery and portal vein mix in the hepatic sinusoid. This mixed blood from the sinusoid then drains into a central vein, which ultimately connects to the hepatic vein and returns to the heart via the inferior vena cava [9]. $C_{Bl,SF}$ and $C_{Bl,SFG}$ are SF and SFG concentrations leaving the blood compartment of the liver (i.e., *outflow* from the liver sinusoids). Within the hepatic sinusoid, blood is in contact with the basolateral membrane allowing for transport of SF and SFG into the hepatocyte via the OATP1B2 transporter [5] and out of the hepatocyte likely via MRP3 transporter [10]. We made two key assumptions for the purposes of model parsimony given limited data: (1) We assumed that all transport kinetic parameters (i.e., $V_{max}$ and $K_m$) were the same for both SF and SFG. (2) The transport flux into the hepatocyte via OATP1B2 and out of the hepatocyte via MRP3 are coupled acting like one single transporter allowing for lumping into one equation with a single, effective $V_{max}$ and $K_m$ (see (9) and (10)). Once inside the hepatocyte, SF has three fates, it can: (1) return to the bloodstream, (2) be metabolized to SFG, and (3) be transported into the bile. Return to the bloodstream and transport into the bile are the two fates for SFG in the hepatocyte. In (11) and (12), $C_{H,SF}$ and $C_{H,SFG}$ represent SF and SFG concentrations contained within hepatocytes. Transport of SF and SFG into the bile is mediated by the transporter MRP2 [5]. It is both the number of transporters ($[E]$) and the transport efficiency per transporter ($k_{cat}$) that determine the overall transport activity (i.e., $V_{max} = [E]k_{cat}$). The bile is drained via catheterization of the common bile duct leading to clearance of both SF and SFG. In (13) and (14), $C_{B,SF}$ and $C_{B,SFG}$ represent SF and SFG bile concentrations. Other notable physiologic parameters are as follows: $V_i$ represents the physiologic volume of each compartment. $K_{m,i}$ represents the Michaelis-Menten constant for each transporter or enzymatic reaction described in Fig. 2. While $K_{m,1}$ has been determined in [11], the other $K_m$ values are not known and were fixed as follows. We assumed the value of $K_{m,2}$ such that it resulted in pseudo-first order kinetics for the SF glucuronidation reaction, and the value of $K_{m,3}$ such that it resulted in an effective zero-order transport of SF and SFG via the MRP2 transporter (i.e., operating at maximal activity). While the $K_m$ of MRP2 for SF is unknown, the $K_m$ for another high affinity substrate is known [12]. To ensure high affinity, we fixed $K_{m,3}$ to be approximately ½ log smaller than this value. Fixing the Michaelis-Menten constants also enables breaking of the correlation between $K_m$ and $V_{max}$ parameters and reduction in the number of unknown parameters to improve confidence in their estimated values. The flow value $F_B$ was fixed as the average empirically determined bile flow rate [13, 14]. Using pseudo-Monte Carlo parameter estimation [15] and fitting to the bile data in concentration units (see section *A*), we estimated the values of $V_{max,1}$, $V_{max,2}$, and $V_{max,3}$, which represent the maximal kinetic efficiencies for the OATP1B2/MRP3 transporter (blood-hepatocyte), the glucuronidation reaction (SF→SFG, [7]), and the MRP2 transporter (hepatocyte-bile), respectively.

## III. RESULTS

*A. SF Fluorescence Measurements Can Be Converted to SF Concentration Measurements Using Nonlinear Regression*

Fig. 3 shows the results of our calibration experiments and modeling using (1)-(3). Estimated parameter values for these empirical equations can be found in Table I.

**TABLE I.** Calibration Curve Parameter Values

| Parameter | Value | Units |
|---|---|---|
| $Int$ | -145.80 | A.U. |
| $Sl$ | 7.77 x 10$^5$ | A.U. mL/mg |
| $I_{max,Pl}$ | 3.99 x 10$^4$ | A.U. |
| $C_{50,Pl}$ | 2.22 x 10$^{-2}$ | mg/mL |
| $I_{max,B}$ | 2.64 x 10$^4$ | A.U. |
| $C_{50,B}$ | 4.63 x 10$^{-2}$ | mg/min |

Fig. 3A is a reproduction of SF fluorescence data from [5] as mean and standard error of the mean (SEM). Fig. 3B shows the results of the *post-hoc* calibration experiments and the linear and nonlinear regression used to relate SF fluorescence measurements to SF concentrations. Fig. 3C shows the results of converting the SF fluorescence data from Fig. 3A to SF concentration units by applying the calibration curves (1)-(3) fitted to data in Fig. 3B.

*B. Empirical Modeling of Hepatic Input for SF and SFG Requires a Concentration-Dependent Partition Coefficient Between Blood and Plasma Compartments*

To begin modeling SF bile disposition, we empirically modeled the input of SF and its metabolite SFG into the liver via summed inflow of the hepatic artery and portal vein. First, we obtained the parameters for (4) by fitting it to the SF blood concentration data in Fig. 3C (Fig. 4A, red curve). The two parts of (4) represent: 1). SF administration and distribution into the blood compartment and 2). A three exponential term accounting for clearance of SF via the kidneys and liver and redistribution into other body compartments. Once an acceptable fit was achieved, we then used (6) to derive the concentration of SF in the plasma. We noticed that a static partition coefficient ($\lambda$), was insufficient to obtain an adequate fit throughout the time course. Thus, we developed a variable partition coefficient described by (5). We then fit the output of (4)-(6) to the plasma SF concentration data (Fig. 4A, blue curve) to estimate parameters for (5) (Fig 4A, inset). Finally, we derived the amount of SF contained within the RBC fraction of the blood (Fig. 4A, cyan curve), which was predictably a very small fraction of SF in the blood. Immediately after SF injection into jugular vein, the SFG concentration in the blood is 0 mg/mL since the dye administered is 2 mg/kg in 0.9% saline and the glucuronidation reaction occurs exclusively in hepatocytes [7]. However, after hepatic metabolism some SFG deposited in the blood will return to the liver. The fraction of SF and SFG in the plasma was determined in humans [8]. We have reproduced and fitted curves to these data for rats in Fig. 4B, by accounting for appropriate vascular transit time differences between humans and rats ($tt_{human} \approx 4\ tt_{rat}$). Using this information and (4)-(6) and (8), we then derived the amount of SFG in the blood and plasma (Fig. 4C). Together, the empirical inflow functions for SF and SFG provided the input for our ODE-based, liver-centric PBPK model. Estimated parameters for (4)-(6) and (8) can be found in Table II.

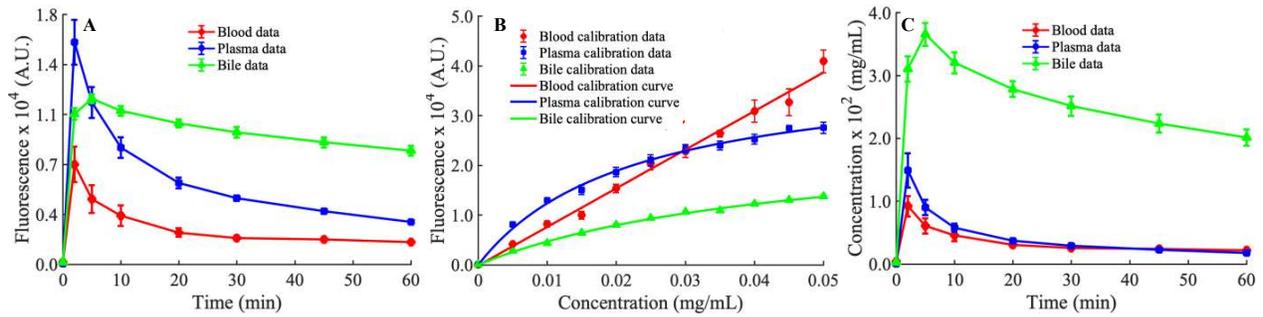

**Fig. 3:** Data from [5]. **A.** Measured blood, plasma, and bile SF fluorescence, shown as mean ± SEM ($N$ = 5). **B.** Calibration curves relating concentration of SF to fluorescence of SF in each of the three regions, shown as mean ± SEM ($N$ = 2-3). **C.** Fluorescent data from panel A are converted to concentration units using the calibration curves from panel B. Data shown as mean ± SEM ($N$ = 5).

**TABLE II.** Empiric SF and SFG Input Parameter Values

| Parameter | Value | Units |
|---|---|---|
| $A_1$ | $2.5 \times 10^{-2}$ | mg/mL |
| $k_1$ | 1.36 | min$^{-1}$ |
| $A_2$ | $6.26 \times 10^{-3}$ | mg/mL |
| $k_2$ | 0.12 | min$^{-1}$ |
| $A_3$ | $2.75 \times 10^{-3}$ | mg/mL |
| $k_3$ | $3.77 \times 10^{-3}$ | min$^{-1}$ |
| $\lambda_{max}$ | 9.00 | Unitless |
| $n_\lambda$ | 3.00 | Unitless |
| $K_\lambda$ | $6.10 \times 10^{-3}$ | mg/mL |
| $f_{max}$ | 0.95 | min$^{-1}$ |
| $n_{SFG}$ | 1.9 | Unitless |
| $t_{50,SFG}$ | 10.03 | min |

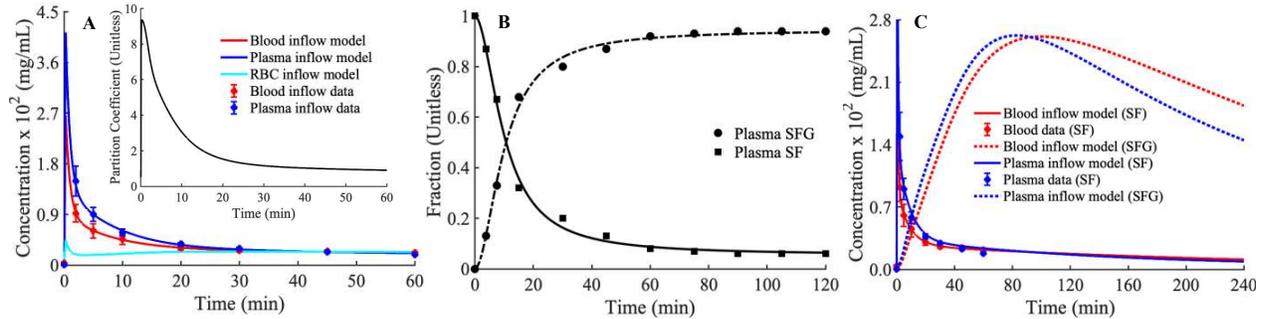

**Fig. 4.** Empirical modeling of whole blood SF fluorescent data. **A.** Fitting of (4) (red curve) to whole blood SF concentration data (red markers, Fig. 2C). Inset is a time-dependent and whole blood SF concentration-dependent partition coefficient (5) allowing for determination of SF concentrations in separate plasma and red blood cell (RBC) compartments (blue and cyan curves). **B.** Fraction of SF and its metabolite SF glucuronide (SFG) in *rat* plasma as determined by [8] in human scaled by ¼ to account for differences in transit time (black markers). Sigmoidal equations were fit to these data to determine time-dependent fraction of SF and SFG. **C.** Empirical input functions for SF from panel A and for SFG derived from panels A and B in whole blood and plasma shown for an extended time period.

*C. Fitting of "Liver-Centric" PBPK Model to Bile SF Concentration Data*

Given the empirical functions for SF and SFG inflow into the liver described in section *B*, we then used these functions as input for the "liver-centric" PBPK model schematized in Fig. 2. This model is described using (9)-(14), and fitting of this model to the bile SF data (Fig. 3C) is shown in Fig. 5. Parameters for this model can be found in Table III. Fig. 5A shows simulated SF and SFG blood and plasma outflow concentrations, while Fig. 5B shows SF bile concentration fit to data and simulated bile SFG and hepatocyte SF and SFG concentrations. Fig. 5C shows simulations for an extended duration demonstrating clearance of SF and SFG over 18h.

*D. Simulations Predict High Sensitivity of bile SF and SFG concentrations to a change in MRP2 Activity*

Intracellular sequestration of MRP2 (i.e., decrease in activity) is hypothesized to be one consequence of IRI leading to a diminished concentration of SF in the bile and hence a smaller SF BPR [5]. To test this hypothesis, we simulated the effects of decreasing the value of each of the estimated parameters ($V_{max,1}$, $V_{max,2}$, and $V_{max,3}$) and analyzing their effects on the amount of SF predicted to be in the bile. Results of these simulations are shown in Fig. 6. Figs. 6A and 6B show the simulated effects of decreasing SF and SFG entry into hepatocytes and decreased conversion of SF to SFG, respectively. Overall, decreased entry into hepatocytes ($V_{max,1}$) had little effect on the amount of SF in the bile. This is speculatively due to the fact that SF uptake into the liver is limited by the flow (i.e., flow-limited) since $V_{max,1}/K_{m,1}$ is large relative to $F$. Diminished conversion of SF to SFG increased the amount of SF in the bile (Fig. 6B). This result is expected since the enzymatic conversion of SF to SFG ($V_{max,2}$) is diminished and a higher concentration of SF is present to be transported into the bile. This trend, however, is opposite to the effect seen in [5]. Finally, simulating a decrease in $V_{max,3}$ (Fig. 6C) results in a large decrease in SF and SFG in the bile. As mentioned previously, this transport process is mediated by MRP2. A decrease in active MRP2 located at the hepatocyte apical membrane is thought to be a result of IRI and leads to a smaller amount of SF in the bile due to a decrease in the number of active transporters. Thus, the decrease in bile SF shown in Fig. 6C, with limited increase in plasma SF, is consistent with our hypothesis that MRP2 is an important contributor to bile SF disposition relative to other factors. As further support for our hypothesis, $V_{max,3}/K_{m,3}$ is close in magnitude to $F$ indicating transport-limited behavior.

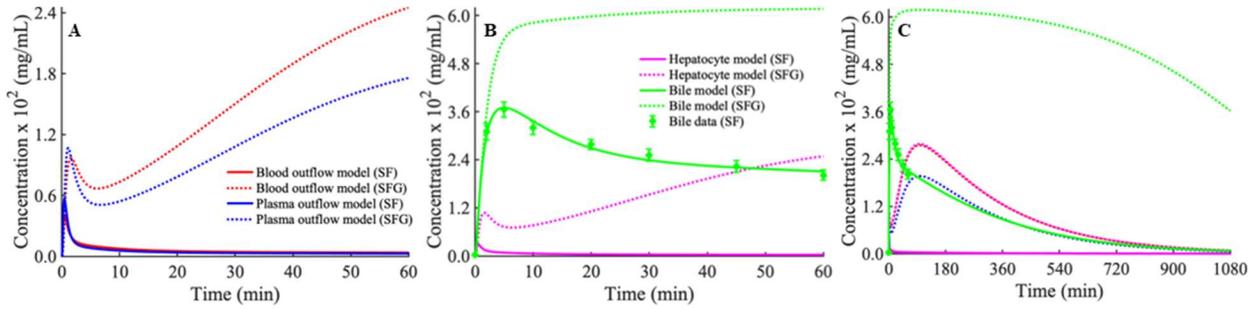

**Fig. 5:** Model parameterization. Fitting of (9)-(14) derived from model schematic in Fig. 2 to bile SF concentration data from Fig. 3C. **A.** Blood and plasma SF and SFG outflow simulations. **B.** Hepatocyte and bile SF and SFG concentrations and bile data. **C.** Model simulations for 18.

**TABLE III.** Liver-Centric PBPK Model Parameter Values

| Parameter | Value | Units |
|---|---|---|
| *Fixed* | | |
| $V_{Bl}$ | 2.01 | mL |
| $F$ | 11.80 | mL/min |
| $K_{m,1}$ | $9.30 \times 10^{-3}$ | mg/mL |
| $V_H$ | 9.16 | mL |
| $K_{m,2}$ | $5.80 \times 10^{-2}$ | mg/mL |
| $K_{m,3}$ | $5.80 \times 10^{-4}$ | mg/mL |
| $V_B$ | $4.47 \times 10^{-2}$ | mL |
| $F_B$ | 0.02 | mL/min$^{-1}$ |
| *Estimated* | | |
| $V_{max,1}$ | 3.12 | mg/min |
| $V_{max,2}$ | 4.29 | mg/min |
| $V_{max,3}$ | $1.47 \times 10^{-3}$ | mg/min |

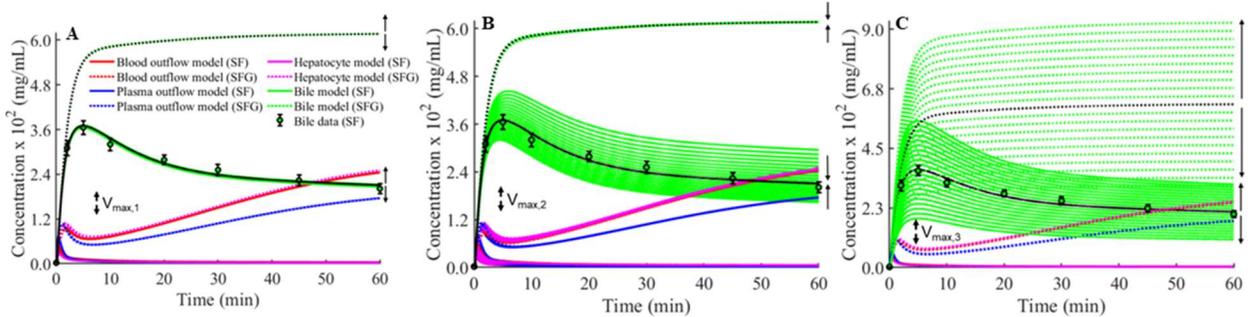

**Fig. 6:** Model predictions with decreased SF transport and metabolism simulating IRI conditions. Arrows next to parameters indicate the direction of change while arrows next to the plots indicate model response. **A.** Simulations showing the effect of decreasing $V_{max,1}$ (relating to transport of SF and SFG into and out of hepatocytes) from 150-50% of original value. **B.** Simulations showing the effect of decreasing $V_{max,2}$ (relating to conversion of SF to SFG inside hepatocytes) from 150-50% of original value. **C.** Simulations showing the effect of decreasing $V_{max,3}$ (relating to transport of SF and SFG into bile) from 150-50% of original value. Changing this value shows the largest effect and is related to the activity of MRP2 as postulated by [5].

## IV. DISCUSSION

In this manuscript, we hypothesized that MRP2 is an important contributor to SF disposition in the bile and that its diminishment markedly decreases the amount of SF in the bile relative to other factors that contribute to SF disposition on passage through the liver. We tested this hypothesis using a computational model of SF liver disposition. Model simulations show that SF concentration in bile is very sensitive to a decrease in the value of the model parameter ($V_{max,3}$) descriptive of MRP2 activity. This suggests that SF along with computational modeling can be used to assess the effect of IRI on MRP2 activity. To the best of our knowledge, this proposed computational model is the first for SF uptake and clearance though the liver. Furthermore, our simulations suggest an optimal protocol for determination of hepatic biliary function using SF. First, while SF shows a decrease in response to a decreased MRP2, SFG shows a much larger range when $V_{max,3}$ is decreased. Thus, for experiments requiring improved sensitivity, probing for SFG may be beneficial. A caveat with this method, however, is that SFG is much less detectable than SF via fluorescence [7], especially in a homogeneous mixture of SF and SFG, likely requiring an advanced separation technique such as HPLC or LC-MS for detection. This method, while capitalizing on sensitivity, may be too slow or costly for clinical use and may be better suited for experiments seeking to validate the model predications presented in this manuscript. Therefore, measuring SF via fluorometry may be an acceptable substitute for clinical use. Also, given that the largest change in SF is in the earlier time points, we recommend that, if SF is to be measured, to limit the duration of the experiment and to emphasize early time points.

While the results presented in this manuscript provide a robust analysis of SF liver uptake and clearance, our approach has some limitations. First, our model relies on empirical input functions to drive SF clearance through the bile. We focused primarily on the hepatic physiology; however, the liver is in the context of many interrelated physiological systems. Thus, future work can expand on our model by developing a "whole-body" model of SF clearance to account for processes such as renal SF clearance. Next, it is known that there are two additional, non-fluorescent, SF metabolites [7]. While we did not account for these in our model, a "whole-body" model would need to consider these metabolites. Finally, we made several assumptions conflating rat and human physiology (e.g., Fig. 4B). These assumptions are unavoidable for the discovery of novel diagnostic and therapeutic strategies. Future studies may focus on obtaining and using human-based SF clearance data.

Once a trend has been established, the ideas regarding SF clearance presented in this manuscript could be applied to clinical practice to help transplant surgeons better assess liver damage due to IRI during the transplant process. This will not only improve outcomes for individual liver transplant recipients, allowing physicians to tailor post-surgical therapy to the individual patient, but will also increase the number of patients able to receive livers that may have otherwise been discarded due to an incomplete understanding of the amount of IRI present in the transplanted liver.

## V. ACKNOWLEDGEMENTS

We thank Dr. Scott S. Terhune for his helpful discussions regarding the importance of this work. All MATLAB code can be found at https://github.com/MCWComputationalBiologyLab/Monti_2023_IEEE.